\documentclass[sigconf]{acmart}

\renewcommand\footnotetextcopyrightpermission[1]{} % removes footnote with conference information in first column
\AtBeginDocument{%
  }

\setcopyright{acmlicensed}
\copyrightyear{2018}
\acmYear{2018}
\acmDOI{XXXXXXX.XXXXXXX}
\acmConference[Conference acronym 'XX]{Make sure to enter the correct
  conference title from your rights confirmation email}{June 03--05,
  2018}{Woodstock, NY}
\acmISBN{978-1-4503-XXXX-X/2018/06}

\usepackage{appendix}
\usepackage{multirow} 
\usepackage{algorithm}
\usepackage{subcaption}
\usepackage{algorithmic}
\usepackage{enumitem}
\usepackage[table]{xcolor} % 必须加 [table] 选项以支持 \cellcolor

\begin{document}

\title{LaRec: Unleashing LLM-based Latent Reasoning for Generative Recommendation}

\author{Yu Xia}
\authornote{Both authors contributed equally to this research.}
\orcid{0009-0002-4128-6968}
\affiliation{%
  % \institution{Institute of Software, Chinese Academy of Sciences }
  \institution{University of Chinese Academy of Sciences}
  \city{Beijing}
  \country{China}
}
\email{xiayu24@mails.ucas.ac.cn}

\author{Zihan Lin}
\authornotemark[1]
\affiliation{%
  \institution{Xiaohongshu}
  \city{Beijing}
  \country{China}
}
\email{linzihan2@xiaohongshu.com}

\author{Wei	Yang}
\affiliation{%
  \institution{Xiaohongshu}
  \city{Beijing}
  \country{China}
}
\email{weiyangvia@gmail.com}

\author{Rui	Zhong}
\affiliation{%
  \institution{Xiaohongshu}
  \city{Beijing}
  \country{China}
}
\email{zhongrui815@gmail.com}

\author{Cheng Chen}
\affiliation{%
  \institution{Xiaohongshu}
  \city{Beijing}
  \country{China}
}
\email{mengde@xiaohongshu.com}

\author{Huan Ren}
\affiliation{%
  \institution{Xiaohongshu}
  \city{Beijing}
  \country{China}
}
\email{renhuan@xiaohongshu.com}

\author{Yao	Hu}
\affiliation{%
  \institution{Xiaohongshu}
  \city{Beijing}
  \country{China}
}
\email{xiahou@xiaohongshu.com}

\renewcommand{\shortauthors}{Trovato et al.}

\begin{abstract}
% Large Language Models (LLMs) have shown great promise in recommendation due to their superior reasoning abilities. However, current methods relying on explicit Chain-of-Thought (CoT) suffer from extreme inference latency caused by autoregressive generation. While Latent Reasoning offers an efficient alternative by reasoning within a continuous latent space, it is hindered by two fundamental challenges: (1) Sparsity of Process Supervision, where sparse feedback from final outcomes fails to guide the evolution of multi-step latent states; and (2) Determinism Constraint, which limits the exploration of diverse user interests and causes misalignment between semantic and collaborative spaces. To overcome these challenges, we propose \textbf{$LaRec$}, a framework designed to unleash the latent reasoning potential of LLMs for generative recommendation. $LaRec$ operates in two stages: First, we design a Latent Pre-training mechanism to endow the LLM with latent reasoning capabilities by incorporating "step-level" and "progressive" alignments as dense supervision. Second, we introduce a Personalized Distribution-Guided Exploration mechanism that utilizes domain knowledge to break determinism and aligns reasoning paths with recommendation goals via Group Relative Policy Optimization (GRPO) algorithm. Experiments on multiple datasets show that $LaRec$ significantly surpasses existing baselines while achieving a substantial improvements in inference efficiency.
Large Language Models (LLMs) have shown great promise in recommendation due to superior reasoning abilities. However, existing methods mainly rely on explicit Chain-of-Thought (CoT), resulting in verbose reasoning texts and inefficient response times. latent reasoning aims to balance efficiency by thinking within a continuous latent space, yet it faces two major challenges: (1) Lack of Fine-grained Supervision: Latent reasoning relies solely on feedback from the final labels, providing sparse supervisory signals that struggle to effectively guide the optimization of multiple hidden reasoning steps. (2) Single Reasoning Path: The deterministic nature of latent reasoning impedes the exploration of users' diverse interests and preferences, thereby limiting the recommendation capabilities of LLMs. To address these issues, we propose \textbf{$LaRec$}, an efficient generative recommendation framework designed to unleash the potential of latent reasoning in LLMs. $LaRec$ consists of two core stages: First, we design Latent Pre-training that empowers LLMs with latent reasoning capabilities by providing rich supervisory signals to the latent space reasoning via step-level alignment and process direction alignment. Second, we introduce Personalized RL-tuning. Specifically, we construct a personalized Gaussian Mixture Distribution for each user based on their historical interests. By randomly sampling distinct reasoning starting points from this distribution during training, we guide the LLMs to traverse diverse reasoning paths within the latent space, enabling efficient exploration of user's multi-faceted interests. Experiments on multiple datasets show that $LaRec$ significantly outperforms existing baselines with comparable efficiency. 
% The  {{\bfseries code\footnote{\url{https://anonymous.4open.science/r/LaRec-DD19}}}} of our implementation has been released. 
\end{abstract}


\maketitle

\section{Introduction}
% Large Language Models (LLMs), empowered by their vast world knowledge and superior logical reasoning capabilities, are reshaping the technical paradigm of recommendation systems (RS) . Unlike traditional methods based on ID matching or shallow semantic matching \cite{sasrec,din,dcn}, LLMs are capable of deeply understanding users' contextual intents and capturing deep interest preferences through multi-step reasoning. To fully unleash this capability, existing works \cite{onerec-think, trackrec, hitlbm} rely on Explicit Chain-of-Thought (Explicit CoT). By employing distillation, supervised fine-tuning, and preference alignment, these methods guide LLMs to analyze user historical behaviors and generate interpretable, step-by-step textual reasoning processes, thereby significantly enhancing both the explainability and accuracy of recommendations. However, Explicit CoT faces a severe efficiency bottleneck in practical applications: inference latency. Due to the token-by-token autoregressive generation mechanism of LLMs, producing a recommendation result with a complete reasoning chain often requires hundreds of network forward passes. This immense computational overhead results in inference latency that struggles to meet the strict real-time requirements of RS.
Large Language Models \cite{gpt,llama, kimi, qwen2.5}, with their extensive world knowledge and logical reasoning capabilities, are driving the evolution of recommender systems from traditional matching paradigms relying on IDs or shallow semantics \cite{DIN, DCN, deepfm, rankmixer, fitmm, yang2025structured} toward deep understanding and reasoning. To fully exploit the reasoning potential of LLMs, existing research \cite{onerec-think, trackrec, hitlbm, r4ec, thinkrec} typically employs explicit CoT, generating step-by-step textual reasoning processes to enhance both interpretability and accuracy of recommendation. However, these approaches introduced a step-by-step generation process with several reasoning tokens, which makes the response latency unaffordable while deploying.

\begin{figure}[h]
\includegraphics[width=0.48\textwidth]{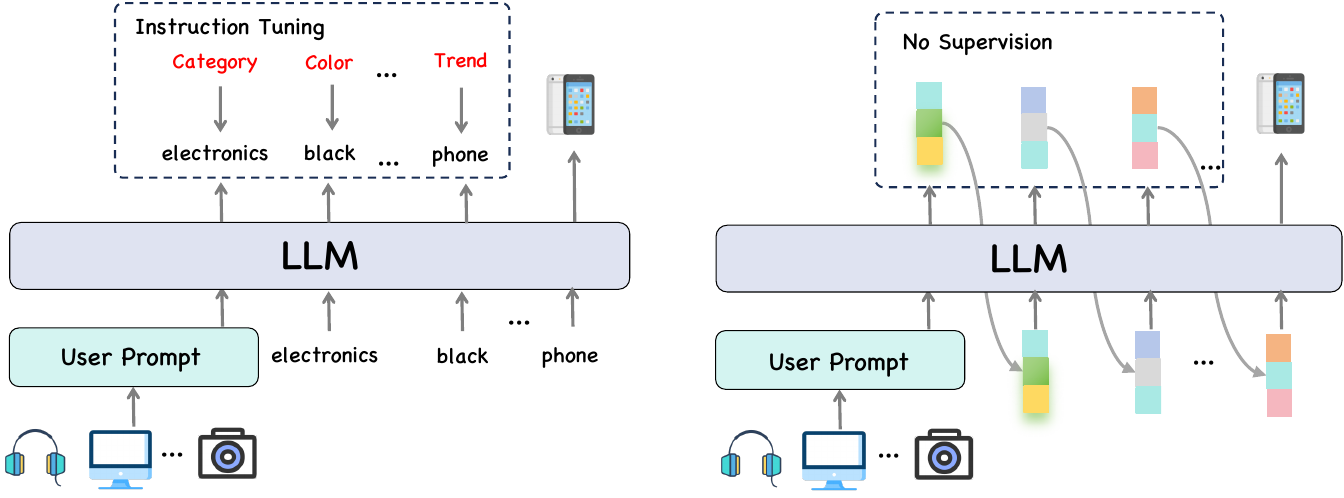}
% \vspace{-9mm}
\caption{Left: Explicit Reason for Recommendation; Right: Latent Reason for Recommendation.}
\label{intro}
% \vspace{-8mm}
\end{figure}
% To mitigate inference efficiency challenges, recent research \cite{coconut, loop_transformer, multiplex} attempts to shift reasoning paths into the latent space of LLMs. Specifically, this approach transforms explicit “step-by-step text generation” into “implicit state evolution” within the model's internal hidden representation space, thereby eliminating the necessity for generating explicit textual CoT. The core advantage of implicit reasoning lies in information density: the high-dimensional vectors of implicit states carry a significantly higher information density than discrete text tokens, implying that concise implicit representations can encode complex reasoning processes. This characteristic enables the model to substantially reduce its dependency on generating lengthy reasoning chains, ultimately achieving more efficient inference.
To improve inference efficiency, recent studies \cite{coconut, loop_transformer, codi, semcot} have begun to explore the path of latent reasoning, which updates prolix text reasoning tokens with several controllable hidden reasoning embeddings. With the help of high-dimensional representations, the latent reasoning approach could encode multiple reasoning steps into limited tokens, thus significantly reducing latency~\cite{simcot}. 

Although latent reasoning theoretically balances effectiveness and efficiency, directly integrating it into recommendation system faces two fundamental challenges in practice: (1) 
\textbf{Lack of Fine-grained Supervision}. As shown in Figure \ref{intro}, the success of Explicit CoT is largely attributed to the loss of Next Token Prediction (NTP) on reasoning tokens at the pre-training stage\cite{gpt,qwen2.5}. Conversely, latent reasoning encapsulates the thinking process into a "black box" where the entire multi-step reasoning chain is optimized solely with the supervision signal from final recommendation target. This sparse and delayed feedback struggles to penetrate deep reasoning steps, failing to effectively guide the learning of intermediate latent states and consequently limiting final recommendation performance. (2) \textbf{Limitations of Deterministic Reasoning Path}. Recommendation tasks inherently require handling the diversity and dynamics of user interests. Explicit CoT can explore diverse reasoning paths through sampling strategies (such as Best-of-N) to identify results that best match the recommendation target. However, current latent reasoning is typically modeled as a deterministic mapping function, allowing the model to execute only a single reasoning trajectory within the latent space \cite{tanthink}. This deterministic nature not only hinders the exploration of diverse user latent interests but also makes it difficult to perform optimization via strategic sampling and alignment, as is done in Explicit CoT. 
% This lack of exploration capability creates an alignment gap between the semantic reasoning space and the recommendation collaborative space, severely capping the potential of LLM. 
%The lack of such exploratory capability prevents LLMs from realizing their reasoning potential,  severely limiting the performance of LLMs in recommendation tasks.
Although some works \cite{latentR3, lares} attempt to introduce exploration by injecting random Gaussian noise during reasoning, several shortages also appear. On one hand, the misalignment between Gaussian noise and reasoning space can lead to a collapse of reasoning process. On the other hand, random exploration could leads to a vast, hard-to-converge search space, making it difficult to achieve effective alignment with recommendation targets.
% To alleviate this issue, some works \cite{latentR3, lares} attempt to introduce exploration by injecting random Gaussian noise into the reasoning process. However, such unstructured stochasticity brings new problems: on the one hand, the semantic misalignment between Gaussian noise and the reasoning space can destabilize or even derail reasoning trajectories; on the other hand, blind random perturbations lead to an exponentially enlarged and poorly structured search space, making effective alignment with recommendation targets extremely difficult.

To address the aforementioned challenges, we propose $LaRec$, an efficient generative recommendation framework. Our design adheres to the principle of "Pre-training then RL Alignment"\cite{r1}.  First, to enhance the supervision during reasoning, we design Latent Pre-training. This stage provides auxiliary supervision signals for each latent state by constructing a step-level alignment with fine-grained preference reasoning signal and process direction alignment between each reasoning step. %enabling LLMs to genuinely learn to "reason within the latent space before making recommendations". 
Specifically, step-level alignment aims to leverage the designability and explainability of explicit reasoning to construct fine-grained guiding signals of user preference reasoning. And distill the organized knowledge from text into the latent reasoning hidden states, to normalize the latent reasoning process and broaden the reasoning space.
Meanwhile, process direction alignment focuses on constraining the reasoning direction between each step, making the latent hidden state progressively point to the final recommendation target to disclose user preference from shadow to deep while reasoning. 

Secondly, we propose Personalized RL-tuning to further strengthen exploration capacity of the LLM itself with personalized guide distribution and reinforcement alignment.
%the break the constraints of deterministic reasoning and avoid blind exploration. 
We construct the personalized Gaussian Mixture Distribution for each user based on their historical interests. During reinforcement alignment,  each reasoning path is independently directed by the user-level distribution at the beginning of reasoning. 
This design constrains the exploration space from unbounded random noise to a user-related semantic space, achieving "anchored exploration". Furthermore, we build our reinforcement alignment with GRPO algorithm \cite{r1}. To precisely align the reasoning process with recommendation targets, we define a composite reward function comprising a sparse hit reward and a dense semantic similarity reward. This design not only provides richer learning feedback but also facilitates more efficient model optimization.

Our contributions are summarized as follows:
\begin{itemize}[leftmargin=*]
\item We propose $LaRec$, an efficient generative recommendation framework. By adhering to the paradigm of "Pre-training then RL Alignment", our framework fully unleashes the implicit reasoning potential of LLMs within recommendation systems.
\item The proposed $LaRec$ incorporates a two-stage training strategy: Latent Pre-training introduces step-level and process-direction alignment, significantly enhancing reasoning coherence and effectiveness; Personalized RL-tuning further optimizes reasoning paths through an efficient "exploration-alignment" mechanism, breaking deterministic limitations and boosting recommendation performance.
\item We conduct extensive experiments on multiple public and industrial datasets to validate the superiority of $LaRec$. Results demonstrate that $LaRec$ not only significantly outperforms existing baselines in recommendation accuracy but also achieves superior inference efficiency.
\end{itemize}

\section{Related Work}
\subsection{Generative Recommendation}
With the rapid advancement of LLMs, the potential of Generative Recommendation (GR) \cite{grsurvey, genrec, hstu, mtgr,forge} has garnered increasing attention. TIGER \cite{tiger} pioneered the generative recommendation paradigm by discretizing item sequences into semantic tokens. OneRec series \cite{onerecv1, onerecv2} explored model architectures, proposing paradigms based on encoder-decoder and decoder-only structures. LC-Rec and HI-Rec \cite{lcrec, hirec} leveraged the natural language understanding capabilities of LLMs to support diverse, task-specific fine-tuning for recommendation tasks. OneRec-Think \cite{onerec-think} empowered LLMs with user understanding capabilities while performing recommendation tasks through a three-stage process involving item alignment, reasoning activation, and reasoning enhancement. Furthermore, several studies focus on optimizing semantic tokens. LETTER \cite{letter} incorporated collaborative filtering embedding vectors and additional loss functions to enhance codebook utilization. DAS \cite{das} proposed a one-stage dual-aligned semantic ID framework that simultaneously optimizes the semantic ID model and the collaborative filtering model, utilizing co-training to reduce information loss during alignment. MMQ \cite{mmq} addressed the issues of modality imbalance and semantic-behavior misalignment by employing multi-expert networks and behavior-aware fine-tuning.

\subsection{Reasoning-Enhanced Recommendation}
The rapid development of test-time compute scaling techniques \cite{r1-like, r-zero} has shifted the research focus in generative AI from Large Language Models (LLMs) \cite{cot, system} to Large Reasoning Models (LRMs) \cite{lrm_survey1, lrm_survey2, lrm_survey3}, injecting new vitality into recommendation systems. Current work on reasoning-enhanced recommendation falls primarily into two categories: (1) Reason-then-Apply. This approach primarily utilizes the powerful reasoning capabilities of LLMs to generate accurate and rich auxiliary information, thereby enhancing the performance of traditional recommendation models. TrackRec \cite{trackrec} continuously guides LLMs to generate more precise user preferences through iterative alternating feedback learning. To utilize user behaviors as comprehensively as possible, HiT-LBM \cite{hitlbm} employs a partition strategy to generate multiple user preferences and combines temporal modeling to yield long-term user representations. (2) Reason-then-Recommend. This approach is mainly applied to end-to-end sequential recommendation, where reasoning is performed prior to recommendation to improve performance. OneRec-Think \cite{onerec-think} conducts explicit user preference analysis before generating the next item. Although this improves recommendation performance, it significantly increases inference latency. Consequently, subsequent work has shifted towards latent reasoning. ReaRec \cite{rearec} is the first to introduce latent reasoning into sequential recommendation. LARES \cite{lares} enhances model representation capabilities by increasing parameter computation density through deep recurrent latent reasoning. $LatentR^3$ \cite{latentR3} applies latent reasoning to LLM-based recommendations. While these methods have achieved certain results, they consistently fail to fully unleash the potential of latent reasoning due to the lack of fine-gained supervision.

\section{Methodology}
\begin{figure*}[h]
\includegraphics[width=0.95 \textwidth]{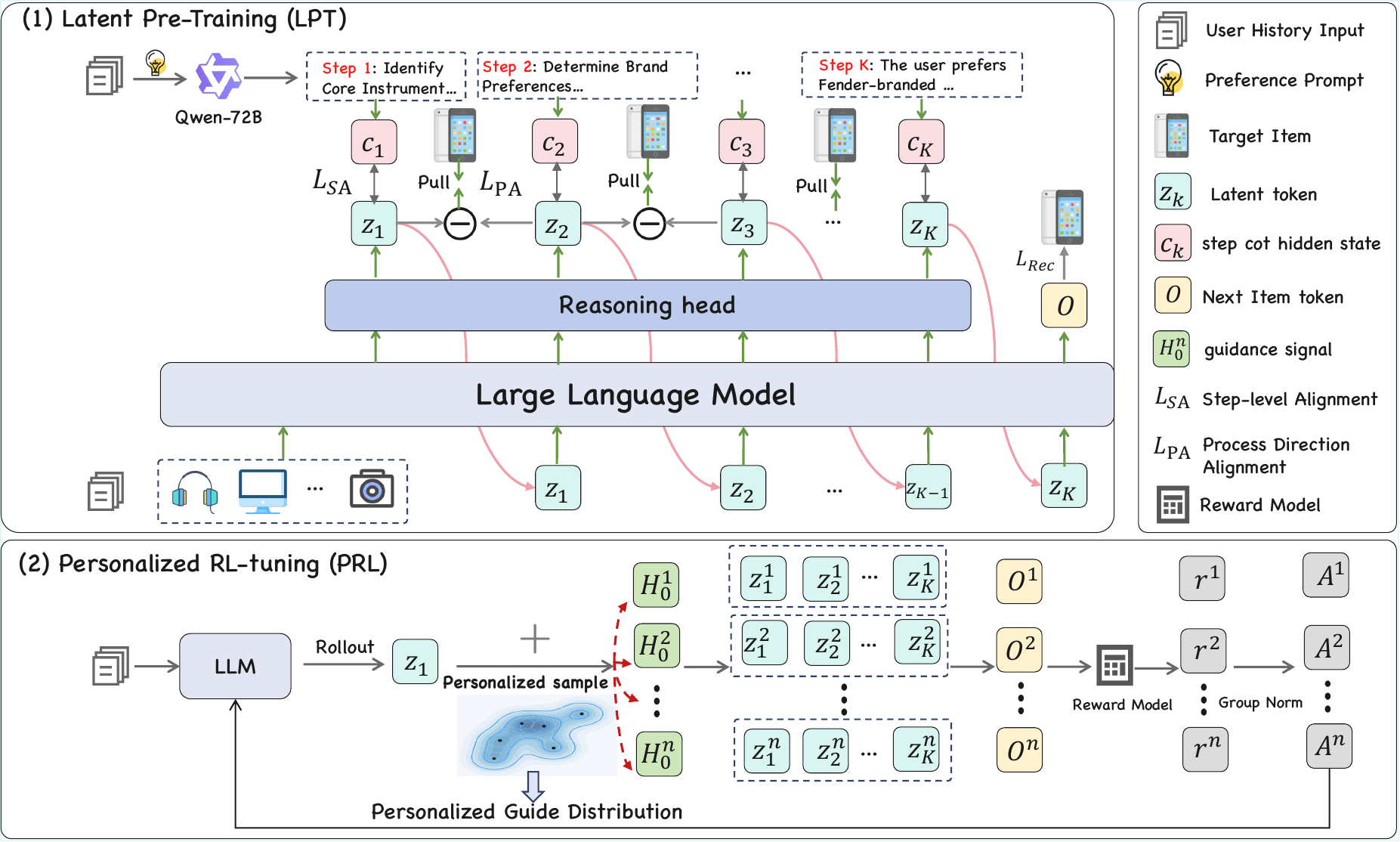}
% \vspace{-3mm}
\caption{An overview of $LaRec$. It mainly consists of two core stages: Latent Pre-Training and Personalized RL-tuning.}
\label{overview}
% \vspace{-5mm}
\end{figure*}
In this section, we first elaborate on the modeling of latent reasoning-enhanced generative recommendation. Subsequently, we detail the core components of $LaRec$, specifically including Latent Pre-training (LPT) for establishing reasoning capabilities and Personalized RL-tuning (PRL) for aligning with recommendation target. The overall framework is illustrated in Figure \ref{overview}.

\subsection{Problem Formulation}
Let $\mathcal{U}$ denote the set of users and $\mathcal{I}$ denote the set of items. For each user $u \in \mathcal{U}$, their historical interaction sequence is represented as $S_u = \{i_1, i_2, \dots, i_t\}$, where each $i_k \in \mathcal{I}$. The goal of generative recommendation is to learn a function $f_{\theta}$ that generates the identifier (ID or Title) of the next item $i_{t+1}$ based on the historical sequence $S_u$. Unlike Explicit CoT methods that generate step-by-step user preferences $C = \{c_1, c_2, \dots , c_n \}$, our objective is to generate a sequence of latent states $Z = \{\mathbf{z}_1, \mathbf{z}_2, \dots, \mathbf{z}_K\}$ within a continuous latent vector space, where $K$ represents the number of reasoning steps. The final recommendation result is predicted based on the reasoning state $\mathbf{z}_{1:K}$:
\begin{equation}
    P(i_{t+1} \mid S_u) = P(i_{t+1} \mid \mathbf{z}_K) \cdot \prod_{k=1}^K P(\mathbf{z}_k \mid \mathbf{z}_{<k}, S_u)
\end{equation}
The standard optimization objective for this task is to maximize the log-likelihood probability of the pair $(S_u, i_{t+1})$. Consequently, the loss for this component, denoted as $\mathcal{L}_{Rec}$, is defined as the standard Cross-Entropy Loss:
\begin{equation}
    \mathcal{L}_{Rec} = - \log P(i_{t+1} \mid S_u,\mathbf{z}_1, \mathbf{z}_2, \dots, \mathbf{z}_K)
\end{equation} 

\subsection{Latent Pre-training}
% The distribution of LLM input embeddings is inconsistent with that of the hidden states in the decoding layer. Consequently, directly utilizing the hidden states from the LLM's decoding layer as inputs causes information disorder, thereby degrading recommendation performance. Inspired by recent works \cite{tanthink, latentR3}, as illustrated in Figure \ref{overview}, we introduce a lightweight Reasoning Head to map the hidden states output by the LLM's decoding layer into the text embedding space, facilitating latent pre-training and alignment. During the Latent Pre-training stage, $LaRec$ learns to perform latent reasoning tailored for recommendation tasks. However, relying solely on the recommendation task loss $\mathcal{L}_{Rec}$ is insufficient to ensure training effectiveness, as this loss fails to provide adequate supervision signals for the intermediate latent reasoning process of $LaRec$. To mitigate this issue, we propose Step-level Alignment and Process Direction Alignment to provide fine-grained supervision signals for the evolution of each latent state.
The effectiveness of explicit CoT in recommendation tasks stems largely from the token-wise, fine-grained supervision introduced during the training phase, which enables the model to learn structured, multi-step reasoning processes. In contrast, latent reasoning occurs entirely within a continuous latent space. If trained solely on the final recommendation task loss $L_{\text{Rec}}$, the intermediate reasoning process lacks direct constraints and is prone to degeneration into shallow pattern matching rather than genuine multi-step reasoning. To address this issue, we propose step-level alignment and process direction alignment mechanisms to provide fine-grained supervision for the evolution of hidden states during latent reasoning. In practice, considering the distribution discrepancy between the LLM's decoder-layer hidden states and the input embedding space, we follow the approach \cite{tanthink, latentR3} by introducing a lightweight Reasoning Head. This head maps latent reasoning states into the text input embedding space to facilitate subsequent alignment and optimization.

\subsubsection{Step-level Alignment }
% Step-level Alignment draws upon the success of Explicit CoT to distill the logical structure of high-quality Explicit CoT into the latent space, thereby facilitating knowledge transfer from explicit text to implicit representations. We employ the explicit CoT as a teacher signal to guide the latent states in learning the teacher's thought trajectory within the semantic space. This process comprises two distinct steps: high-quality CoT construction and alignment optimization.
Step-level Alignment draws upon the successful experience of explicit CoT. It distills the logical structure of explicit, fine-grained user preferences into the latent space, thereby facilitating the knowledge transfer from explicit preferences to implicit representations. Using fine-grained user preferences as teacher signals, we guide the latent states to learn the teacher's reasoning trajectory within the semantic space. This process involves two key steps: Fine-grained User Preference Extraction and Alignment Optimization.

\begin{figure}[h]
\includegraphics[width=0.48\textwidth]{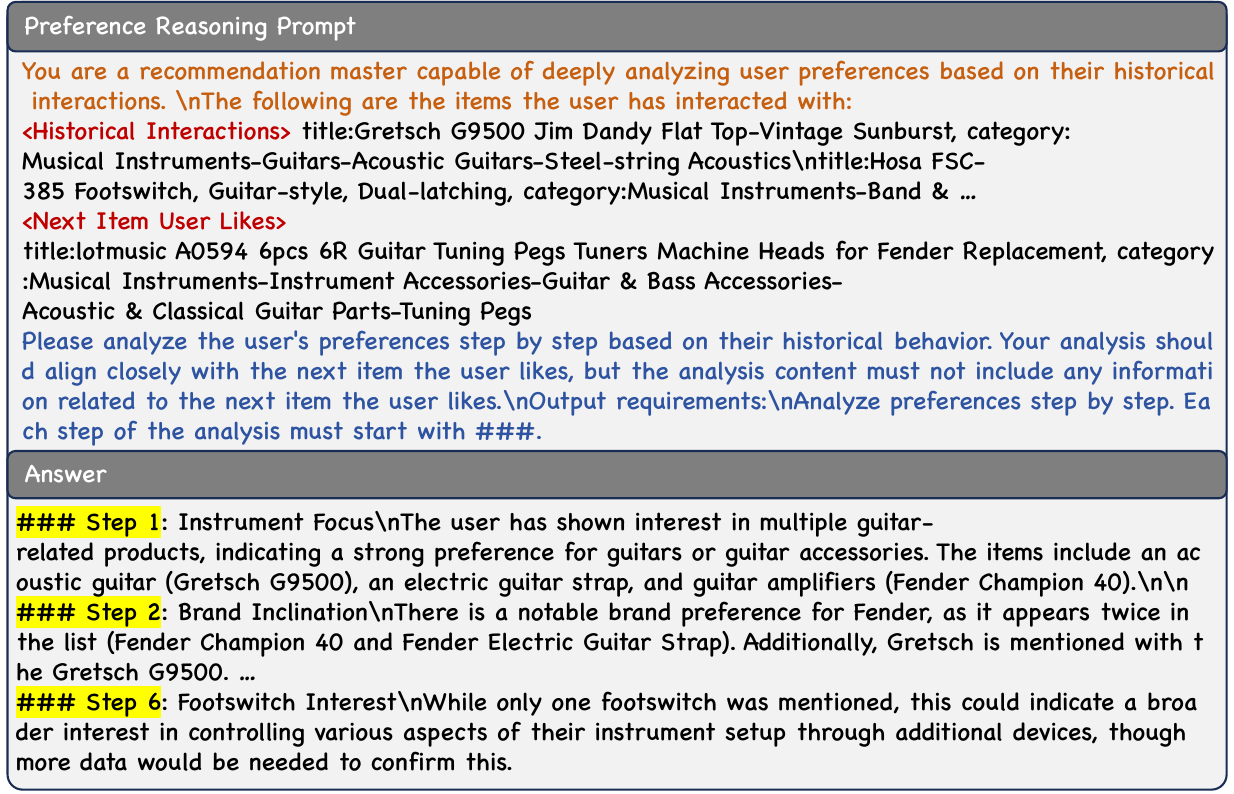}
% \vspace{-7mm}
\caption{Preference Reasoning Prompt: Example prompt and response.}
\label{prompt}
% \vspace{-6mm}
\end{figure}

\paragraph{Fine-grained User Preference Extraction}
To provide accurate guidance for latent reasoning, it is essential to extraction explicit, fine-grained user preference CoT. To this end, we employ a Target-Oriented Reverse Reasoning Generation strategy. Specifically, for each training instance $(S_u, i_{tgt})$, we feed both the user history $S_u$ and the ground-truth target item $i_{tgt}$ into a powerful reasoning model like Qwen-72B \cite{qwen2.5}. We design a user preference reasoning prompt $prompt^{cot}$, as illustrated in Figure \ref{prompt}, which instructs the LLM to generate a logically coherent reasoning path $C^+ = \{c_1, c_2, \dots, c_K\}$. This path must originate from the user history and logically deduce the target item. However, the reasoning process itself can not contain any information regarding the target item $i_{tgt}$. Subsequently, we utilize LLM to extract the hidden states of these CoT steps in the following manner, denoted as $H = \{\mathbf{h}_1, \dots, \mathbf{h}_K\}$:
\begin{equation}
    h_k = \textbf{LLM}_{last}(prompt^{cot}(S_u,i_{tgt}), c_{1:k})
\end{equation}
where $\text{LLM}_{last}(\cdot)$ denotes the output of the LLM’s final decoding layer. The LLM here is the same one used for the latent pre-training mentioned below.
% To enhance computational efficiency, we adopt an In-batch Negatives strategy, utilizing the Hidden States from other samples within the same batch as negative samples $H^-$.

\paragraph{Alignment Optimization}
% We use the constructed fine-grained, high-quality explicit CoT as the guiding signal (i.e., the positive samples $H^+= \{\mathbf{h}_1^+, \dots, \mathbf{h}_K^+\}$). To improve efficiency, we employ an In-batch Negatives strategy, utilizing the hidden states of other samples within the same batch as the negative samples $H^-$.
We represent both the explicit and latent reasoning processes as state sequences composed of multiple steps. For each training sample, we construct a fine-grained, high-quality explicit CoT reasoning trajectory to obtain a sequence of explicit reasoning states $H = \{h_k\}_{k=1}^K$. Simultaneously, during the latent reasoning, the model internally generates a corresponding sequence of hidden states $\{\mathbf{z}_k\}_{k=1}^K$. To establish a step-level correspondence between these two processes, we constrain each latent reasoning state $\mathbf{z}_k$ to maintain semantic consistency with its corresponding explicit reasoning state $\mathbf{h}_k^+$ in the representation space, while ensuring it remains distinguishable from the reasoning states of other samples. Within a mini-batch, the explicit reasoning states of other samples naturally form a contrastive reference set $H^-$, enabling stable state-alignment constraints without the need for additional manual annotation. To this end, we employ an InfoNCE-based contrastive loss for step-level state alignment:
% \begin{equation}
%     \mathcal{L}_{SA} =  \sum_{k=1}^K -\log \frac{\exp(\text{s}(\mathbf{z}_k, \mathbf{h}_k^+) / \tau_1)}{\exp(\text{s}(\mathbf{z}_k, \mathbf{h}_k^+) / \tau_1) + \sum_{j \in \text{B}, j \neq i} \exp(\text{s}(\mathbf{z}_k, \mathbf{h}_{k, j}^-) / \tau_1)}
% \end{equation}
\begin{equation}
    \mathcal{L}_{SA}=\sum_{i=1}^B\sum_{k=1}^K-\log\frac{\exp\left(\text{sim}(\mathbf{z}_{i,k},\mathbf{h}_{i,k})/\tau_1\right)}{\sum_{j=1}^B\exp\left(\text{sim}(\mathbf{z}_{i,k},\mathbf{h}_{j,k})/\tau_1\right)}
\end{equation}
% \begin{equation}
%     \mathcal{L}_{SA}=\sum_{i=1}^B\sum_{k=1}^K-\log\frac{\exp\left(\sin(\mathbf{z}_{i,k},\mathbf{h}_{i,k}^+)/\tau_1\right)}{\exp\left(\sin(\mathbf{z}_{i,k},\mathbf{h}_{i,k}^+)/\tau_1\right)+\sum_{j=1,j\neq i}^B\exp\left(\sin(\mathbf{z}_{i,k},\mathbf{h}_{j,k}^-)/\tau_1\right)}
% \end{equation}
where $\text{sim}(\cdot)$ denotes cosine similarity and $\tau_1$ is the temperature coefficient. This loss constrains the representation consistency between latent and explicit reasoning states at the step level. It encourages LLMs's latent reasoning process to progressively approximate the semantic structure of the explicit CoT reasoning trajectory, thereby mitigating the issue of reasoning degeneration to a certain extent.

\subsubsection{Process Direction Alignment}
While Step-level Alignment ensures the semantic validity of individual reasoning steps, it fails to guarantee that these steps are coherently directed towards the final target within the high-dimensional embedding space. In the absence of directional constraints, the evolution of latent states is prone to "semantic idling" \cite{simcot}—a phenomenon where the model stagnates locally within the semantic space, thereby degrading recommendation performance. To address this, Process Direction Alignment leverages the target item as a directional guide. Specifically, for a target item $i_{tgt}$, we apply Mean Pooling to the embeddings of its title text tokens to derive the target semantic vector $\mathbf{e}_{tgt}$:
\begin{equation}
\mathbf{e}_{tgt} = \text{MeanPooling}({LLM_{TKZ}}(i_{tgt}))
\end{equation}
where $LLM_{TKZ}$(·) presents the LLM tokenizer and word embedding layer, encapsulating the process of transforming textual metadata into token representations.

Building upon this, we define the actual evolutionary direction of the $({k+1})^{th}$ latent reasoning step as $\mathbf{v}_{k+1} = \mathbf{z}_{k+1} - \mathbf{z}_k$. Drawing on the principle of "linear accumulation" \cite{111}, each reasoning step should function as a sub-component of the final target $\mathbf{e}_{tgt}$. Consequently, the ideal update vector corresponds to the direction of the target item itself, scaled by a factor of  $\frac{1}{K-1}$. We constrain the reasoning process by maximizing the cosine consistency between the actual update vector $\mathbf{v}_{k+1}$ and the scaled target vector:
\begin{equation}
    \mathcal{L}_{PA} = \sum_{k=2}^{K} \max\left(0, 1 - \text{sim}\left(\mathbf{v}_k, \frac{1}{K-1} \mathbf{e}_{tgt}\right)\right)
\end{equation}
where $\text{sim}(\cdot)$ denotes cosine similarity, and $K$ represents the maximum steps of latent reasoning.

\subsubsection{Overall Optimization}
The overall objective function for Latent Pre-training is formulated as:
\begin{equation}
    \mathcal{L}_{LPT} = \mathcal{L}_{Rec} + \alpha \mathcal{L}_{SA} + \gamma \mathcal{L}_{PA}
\end{equation}
where $\alpha$ and $\gamma$ are hyperparameters designed to balance the weights of the different loss components during the optimization process. Upon completion of Latent Pre-training, we obtain the pre-trained model, denoted as $LLM_{pretrain}$.

\subsection{Personalized RL-tuning}
% Following Latent Pre-training, the model $LLM_{pretrain}$ possesses latent reasoning capabilities. However, this process remains inherently a deterministic mapping. To break the constraints of deterministic reasoning while avoiding blind exploration, this stage introduces the Personalized Reinforcement Alignment mechanism. Specifically, we leverage recommendation domain knowledge to construct a guidance distribution to break determinism, and subsequently align the model with recommendation targets via GRPO\cite{r1}.

After undergoing latent pre-training, the model $LLM_{pretrain}$ acquires implicit reasoning capabilities. However, the entire implicit reasoning process remains essentially a deterministic mapping, which limits the exploration potential of the LLM. To break the deterministic constraint while avoiding blind exploration, this stage introduces a Personalized RL-tuning mechanism. Specifically, we design a personalized behavior distribution tailored to the characteristics of the recommendation domain. This allows the LLM to explore within the latent space and align with recommendation objectives via GRPO\cite{r1}, thereby further enhancing the model's recommendation performance.

\subsubsection{Personalized Guide Distribution Construction}
The conventional approach to breaking determinism involves introducing random noise (e.g., standard Gaussian noise $\mathcal{N}(\mu, \sigma^2)$). However, in a high-dimensional latent space, blind random perturbations are prone to disrupting the semantic structure established during the latent pre-training phase, leading to a collapse of reasoning logic. To achieve "effective exploration", we propose an Anchored Exploration strategy, positing that exploration should originate from the user's interest neighborhood rather than from random Gaussian noise. We leverage the user's historical interaction items as "semantic anchors" for exploration. We define the user's Interest Anchor Set $\Omega_u$ as the set of embeddings corresponding to their historical interactions:
\begin{equation}
    \begin{gathered}
        \mathbf{e}_{i} = \operatorname{MeanPooling}\bigl(\text{LLM}_{\text{TKZ}}(i_{t})\bigr), \\
        \Omega_u = \{ \mathbf{e}_t \mid i_t \in S_u \}
    \end{gathered}
\end{equation}
where $LLM_{TKZ}$(·) presents the LLM tokenizer and word embedding layer, $e_t$ denotes the semantic representation of item $i_t$, and $S_u$ denotes a user's interaction sequence.

We model the guidance signal $H_0$ as a Uniform Gaussian Mixture Model centered on $\Omega_u$. This embodies the assumption that every historical behavior of the user has the potential to serve as a starting point for triggering a new interest:
\begin{equation}
    H_0 \sim \mathcal{P}_{\text{guide}}(u) = \frac{1}{|\Omega_u|} \sum_{\mathbf{e} \in \Omega_u} \mathcal{N}(\mathbf{e}, \sigma^2 \mathbf{I})
\end{equation}
where $\sigma$ is a hyperparameter controlling the local exploration radius. By sampling $H_0$ from $\mathcal{P}_{\text{guide}}(u)$, we introduce the randomness to break determinism while simultaneously ensuring that the perturbation vectors remain within the effective scope of the User Interest Manifold.

\subsubsection{Reinforcement Alignment}
Next, we employ the Personalized Guide Distribution to conduct reinforcement learning, aligning with recommendation objectives and further enhancing the recommendation capabilities of LLMs. 
% The pseudocode of Reinforcement Alignment is detailed in Algorithm \ref{alg:grpo_alignment} in Appendix \ref{A}.
\paragraph{Rollout}
We inject the sampled guidance signal $H_0$ into the reasoning process as a perturbation to the initial latent state. Consequently, the input latent state of the model is modified as 
\begin{equation}
    \mathbf{z}_1^i = \mathbf{z}_1 + H_0^i
\end{equation}
where $H^i_0$ denotes the guidance signal for the $i^{th}$ rollout.

During the training phase, as shown in figure \ref{overview}, for each user $u$, we execute the sampling process as follows: we independently sample from $\mathcal{P}_{\text{guide}}(u)$ $n$ times to generate $n$ parallel latent reasoning paths, yielding a corresponding set of prediction results $\mathcal{O} = \{O^1, O^2, \dots, O^n\}$.

\paragraph{Reward}
To balance recommendation accuracy with semantic relevance, we design a composite reward function $R(O^i, i_{tgt})$ for the latent reasoning paths:
% \begin{equation}
%     R(O_i) = \underbrace{\mathbb{I}(O_i = v_{tgt})}_{\text{Hard Reward (Hit)}} + \beta \cdot \underbrace{\text{sim}(\text{Emb}(O_i), \text{Emb}(v_{tgt}))}_{\text{Soft Reward (Similarity)}}
% \end{equation}
\begin{equation}
    r(O^i) = \mathbb{I}(O^i = i_{tgt}) + \lambda \cdot \text{sim}(\text{Emb}(O^i), \text{Emb}(i_{tgt}))
\end{equation}
where $\mathbb{I}(\cdot)$ is the indicator function, providing a sparse "hit" signal; $\text{sim}(\cdot)$ computes the semantic cosine similarity between the predicted item and the ground-truth item. This latter component provides a dense signal, enabling the model to discern the direction for optimization even in cases of non-hits.

\paragraph{Optimization.}
Based on the designed reward, we use the GRPO algorithm to conduct alignment training for the latent pre-trained $LLM_{pretrain}$. For each user's $n$ reasoning paths and prediction results, we calculate the reward values for each path ($r^1$, $r^2$, ..., $r^n$). Subsequently, we compute the relative advantages of each latent reasoning path within the group to guide the optimization direction:
\begin{equation}
A^i = \frac{r^i - \mu}{\sigma}, 
\quad \mu = \frac{1}{n}\sum_{i=1}^{n} r^i, \quad 
\sigma = \sqrt{\frac{1}{n}\sum_{i=1}^{n}\left(r^i - \mu\right)^2}
\end{equation}
Here, $A^i$ represents the advantage of the $i^{th}$ latent reasoning path within the group, and $\mu$ and $\sigma$ represent the mean and standard deviation, respectively.
Therefore, our GRPO optimization objective is:
\begin{align}
\mathcal{J}_{\text{GRPO}}(\theta) 
&= \mathbb{E}\!\left[ \{ O^i\}_{i=1}^{n} \sim LLM_{\theta_{\text{old}}}(O^i \mid z^i_{1:K}, S_u) \right] \nonumber \\
&\quad \frac{1}{n} \sum_{i=1}^{n} \left( \rho^i - \beta D_{\text{KL}} \left( LLM_{\theta} \parallel LLM_{\text{ref}} \right) \right), \nonumber \\
\rho^i &= \min \left[ \delta^i A^i,\; \text{clip}\!\left(\delta^i, 1 - \epsilon, 1 + \epsilon \right) A^i \right], \nonumber \\
\delta^i &= 
\frac{LLM_{\theta}\!\left(O^i \mid z^i_{1:K}, S_u\right)}
     {LLM_{\theta_{\text{old}}}\!\left(O^i \mid  z^i_{1:K} , S_u\right)}
\end{align}
Here, $LLM_{\theta}$ and  $LLM_{\text{ref}}$ are initialized as $LLM_{pretrain}$, $\beta$ and $\epsilon$ are hyperparameters.

% \subsection{Discussion}

\section{Experiments}
\begin{table}[h]
\centering
\caption{Statistics of datasets used in this paper} % 表格标题
\vspace{-3mm}
\setlength{\tabcolsep}{3pt} % 减小列间距
\begin{tabular}{lcccccc}
\toprule
{Dataset} & {\#Train} & {\#Valid} & {\#Test} & {\#User} & {\#Item} & {Sparsity} \\
\midrule
Toys  & 53898  & 6737  & 6738 & 11311 & 6299 & 99.89\% \\
Instruments & 66500 & 8312 & 8313 & 12044 & 5030 & 99.84\%\\
Movie  & 121009  & 15126  & 15127 & 2025 & 3405 & 99.14\% \\
Industry  & 294624  & 36828  & 36829 & 41414 & 7398 & 99.87\% \\

\bottomrule
\end{tabular}
\label{tab:statistics}
\end{table}

\begin{table*}[h]
\centering
\renewcommand{\arraystretch}{1.1} 
\setlength{\tabcolsep}{3pt} 
\caption{Performance Comparison. The best results are highlighted in \colorbox{green!10}
{\textbf{bold}} and the second-best results are \underline{underlined}.}
\vspace{-3mm}
\label{overall}
\begin{tabular}{llccccccccccc}
\toprule
% \multirow{2}{*}{Datasets} & Methods & \multicolumn{4}{c}{ID based} & \multicolumn{5}{c}{LLM based} & \multirow{2}{*}{Rel.Impr.} \\ 
\multirow{2}{*}{Datasets} & Methods & \multicolumn{4}{c}{ID based} & \multicolumn{6}{c}{LLM based}  \\
\cmidrule(lr){3-7} \cmidrule(lr){8-13}
 & Metrics & GRU4Rec & SASRec & BERT4Rec & ReaRec & PLR & TallRec & BIGRec & $D^3$ & TrackRec & $LatentR^3$ & \cellcolor{green!10}$LaRec$ \\ 
\midrule

\multirow{4}{*}{Toys} 
 & H@5  & 0.0417 & 0.0700 & 0.0709 & 0.0721 & 0.0752& 0.0651 & 0.0560& 0.0651 & 0.0693 & \underline{0.0785} & \cellcolor{green!10}\textbf{0.0813\textsuperscript{*}} \\
 & H@10 & 0.0564 & 0.0884 & 0.0888 & 0.0900 & 0.0933 & 0.0837 & 0.0802 & 0.0900 & 0.0969 & \underline{0.1055} & \cellcolor{green!10}\textbf{0.1073\textsuperscript{*}} \\
 & N@5  & 0.0305 & 0.0427 & 0.0438 & 0.0517 & 0.0540 & 0.0479 & 0.0420 & 0.0499 & 0.0496 & \underline{0.0573} & \cellcolor{green!10}\textbf{0.0612\textsuperscript{*}} \\
 & N@10 & 0.0352 & 0.0487 & 0.0496 & 0.0575 & 0.0590 & 0.0539 & 0.0497 & 0.0566 & 0.0586 & \underline{0.0657} & \cellcolor{green!10}\textbf{0.0695\textsuperscript{*}} \\
\midrule

\multirow{4}{*}{Instruments} 
 & H@5  & 0.0834 & 0.0762 & 0.0756 & 0.0897 & 0.0968 & 0.0926 & 0.0920 & 0.0938 & 0.0959 & \underline{0.0997} & \cellcolor{green!10}\textbf{0.1024\textsuperscript{*}}  \\
 & H@10 & 0.0949 & 0.0850 & 0.0852 & 0.1065 & 0.1175 & 0.1082 & 0.1093 & 0.1106 & 0.1180 & \underline{0.1208} & \cellcolor{green!10}\textbf{0.1238\textsuperscript{*}}  \\
 & N@5  & 0.0709 & 0.0676 & 0.0658 & 0.0778 & 0.0847 & 0.0819 & 0.0783 & 0.0839 & 0.0793 & \underline{0.0877} & \cellcolor{green!10}\textbf{0.0894\textsuperscript{*}}  \\
 & N@10 & 0.0746 & 0.0704 & 0.0689 & 0.0832 & 0.0913 & 0.0870 & 0.0860 & 0.0893 & 0.0864 & \underline{0.0939} & \cellcolor{green!10}\textbf{0.0963\textsuperscript{*}}  \\
\midrule

\multirow{4}{*}{Movie}    
 & H@5  & 0.1153 & 0.1431 & 0.1349 & 0.1409 & 0.1468 & 0.1526 & 0.1512 & \underline{0.1535} & 0.1471 & 0.1528 & \cellcolor{green!10}\textbf{0.1648\textsuperscript{*}}  \\
 & H@10 & 0.1661 & 0.1849 & 0.1801 & 0.2156 & 0.2168 & 0.2172 & 0.2160 & 0.2083 & 0.1840 & \underline{0.2175} & \cellcolor{green!10}\textbf{0.2256\textsuperscript{*}}  \\
 & N@5  & 0.0802 & 0.1037 & 0.0975 & 0.0868 & 0.0888 & 0.1075 &0.1068 & 0.1104 & \underline{0.1116} & 0.1078 & \cellcolor{green!10}\textbf{0.1174\textsuperscript{*}}  \\
 & N@10 & 0.0965 & 0.1171 & 0.1121 & 0.1109 & 0.1123 & 0.1284 &0.1268 & 0.1281 & \underline{0.1335} & 0.1281 & \cellcolor{green!10}\textbf{0.1370\textsuperscript{*}}  \\
\midrule

\multirow{4}{*}{Industry}     
 & H@5  & 0.1706 & 0.1956 & 0.1897 & 0.2485 & 0.2612 & 0.2628 &0.2540 & 0.2880 & \underline{0.2900} & 0.2899 & \cellcolor{green!10}\textbf{0.3033\textsuperscript{*}}  \\
 & H@10 & 0.2612 & 0.2914 & 0.2847 & 0.3285 & 0.3436 & 0.3654 &0.3462 & 0.4010 & 0.3827 & \underline{0.4020} & \cellcolor{green!10}\textbf{0.4090\textsuperscript{*}}  \\
 & N@5  & 0.1116 & 0.1313 & 0.1284 & 0.1768 & 0.1955 & 0.1842 & 0.1798 & 0.2029 & 0.2071 & \underline{0.2079} & \cellcolor{green!10}\textbf{0.2168\textsuperscript{*}}  \\
 & N@10 & 0.1407 & 0.1622 & 0.1590 & 0.2027 & 0.2187 & 0.2174 & 0.2045 & 0.2394 & 0.2371 & \underline{0.2425} & \cellcolor{green!10}\textbf{0.2509\textsuperscript{*}}  \\ 
\bottomrule
\multicolumn{13}{l}{* denotes statistically significant improvement (measured by t-test with p-value<0.001) over baselines.}
\end{tabular}
\end{table*}

\subsection{Experiment Setup}
\subsubsection{Datasets}
Experiments are conducted based on the public Amazon datasets and MovieLens-100K. For the Amazon datasets, we evaluate the $LaRec$ primarily on two specific domains: Toys and Instruments. Additionally, we construct a real-world offline industrial dataset to assess the model's robustness under realistic data distribution conditions. Following previous work \cite{latentR3}, we applied a 5-core filtering strategy to exclude inactive users and unpopular items with fewer than 5 interactions, thereby ensuring the reliability of the evaluation results. To simulate real-world application scenarios, we perform a temporal split on the preprocessed datasets based on interaction timestamps, partitioning them into training, validation, and testing sets in an 8:1:1 ratio. Furthermore, to align with the requirements of baseline models across all experimental settings, we standardize the maximum length of user interaction sequences to 10, ensuring consistency in sequential pattern analysis. Key statistics of these datasets are summarized in Table \ref{tab:statistics}.

\subsubsection{Baseline Models}
% For the Top-N recommendation task, we compare our proposed $LaRec$ against two categories of baselines. (1) Traditional Sequential Recommendation Models: GRU4Rec \cite{gru4rec} applies RNNs to capture session-based patterns. SASRec \cite{SASRec} and BERT4Rec \cite{bert4rec} utilize unidirectional and bidirectional self-attention mechanisms. ReaRec \cite{rearec} incorporates latent reasoning into sequential modeling. PLR \cite{plr} focuses on ID-based sequential recommendation via parallel aggregation of multiple latent reasoning paths. (2) LLM-based Recommendation Methods: TallRec \cite{tallrec} directly leverages LLMs to predict target items based on user history. BIGRec \cite{bigrec} ranks items by mapping generated semantics to the actual item space via distance metrics. $D^3$ \cite{d3} is a representative generative recommendation method that fine-tunes LLMs to perform next-item prediction. TrackRec \cite{trackrec} employs a "reason-then-recommend" approach, first generating user preference CoT via reasoning before making recommendations based on these preferences. $LatentR^3$ \cite{latentR3} directly applies implicit reasoning to LLM-based generative recommendation. To ensure a fair evaluation, for LLM-based recommendation methods, a Trie-constrained beam search \cite{latentR3} is uniformly employed to generate Top-N items. 
For the Top-N recommendation task, we compare our proposed $LaRec$ against two categories of baselines. (1) Traditional Sequential Recommendation Models: GRU4Rec \cite{gru4rec} captures sequential user behavior patterns via RNNs. SASRec \cite{SASRec} and BERT4Rec \cite{bert4rec} utilize uni- and bi-directional self-attention. ReaRec \cite{rearec} integrates latent reasoning into sequential modeling. PLR \cite{plr} aggregates parallel latent reasoning paths for ID-based recommendation. (2) LLM-based Methods: TallRec \cite{tallrec} directly predicts target items from history. BIGRec \cite{bigrec} ranks items by mapping generated semantics to item space. $D^3$ \cite{d3} is a representative generative recommendation method that fine-tunes LLMs to perform next-item prediction. TrackRec \cite{trackrec} adopts a "reason-then-recommend" paradigm with explicit CoT. $LatentR^3$ \cite{latentR3} applies implicit reasoning to llm based generative recommendation. For a fair comparison, all LLM-based methods uniformly employ Trie-constrained beam search \cite{latentR3} to generate Top-N items.
% below: \textbf{GRU4Rec} utilizes Gated Recurrent Units to capture sequential user behavior patterns. \textbf{SASRec} is a Transformer-based model that encodes user interaction sequences via a unidirectional multi-head self-attention mechanism. \textbf{BERT4Rec} employs a bidirectional self-attention architecture, applying the masked prediction mechanism to sequence modeling. \textbf{ReaRec} introduces multi-step autoregressive computation within an ID-based latent space to explore test-time reasoning mechanisms. \textbf{TallRec} directly leverages LLMs to predict target items based on user history; for this study, we adapt it to generate the next item. \textbf{BIGRec} grounds the recommendation space to the actual item space, computes the distance between generated semantics and real item representations, and obtains a ranking of actual items. \textbf{ D$^3$ } is a representative generative recommendation method that fine-tunes LLMs to perform next-item prediction. Additionally, it incorporates debiasing techniques during the inference phase to enhance the quality of generated recommendations. \textbf{TrackRec} employs a "reason-then-recommend" approach, first generating user preference CoT via reasoning before making recommendations based on these preferences. \textbf{LatentR$^3$} directly applies implicit reasoning to LLM-based generative recommendation.

\subsubsection{Evaluation Metrics}
We evaluate the Top-N recommendation effectiveness using Hit Ratio (HR@N) and Normalized Discounted Cumulative Gain (NDCG@N), with N set to 5 and 10. For brevity, we denote HR@5 and NDCG@5 as H@5 and N@5, respectively, similarly for "@10" cases.

\subsubsection{Implemenation}
For traditional recommendation models, we use the Adam optimizer and perform grid searches over learning rates in \{ $1e^{-3}$, $5e^{-4}$, $1e^{-4}$ \} and weight decay values in \{ $1e^{-4}$, $1e^{-5}$, $1e^{-6}$ \}, and all models are trained using Binary Cross-Entropy loss with randomly sampled negative items. For LLM-based methods, we use Qwen2.5-1.5B Team \cite{qwen2.5} as the backbone LLM. Latent Pre-training is conducted using the AdamW optimizer, with learning rates selected from \{ $ 3e^{-4}$, $5e^{-5}$, $1e^{-5}$ \}, and early stopping is applied with a patience of 1.  Based on the number of steps in the constructed high-quality CoT, we set the latent reasoning step count $K$ = 6. The alignment loss weights $\alpha$ and $\gamma$ in $L_{LPT}$ are both set to 1.0. During the reinforcement learning stage, we set local exploration radius $\sigma$ = 1.5, $\lambda$ = 1.0, and search learning rates within \{ $1e^{-5}$, $1e^{-4}$, 5e$^{-4}$ \}. All experiments are run on 8 NVIDIA H20 GPUs.

\subsection{Overall Performance}

We evaluate $LaRec$ against various baseline methods across three public datasets and one offline industrial dataset. These baselines are categorized into two groups: (1) ID-based sequential recommendation methods, and (2) LLM-based sequential recommendation methods. Table \ref{overall} presents the comprehensive experimental results, from which we derive the following core observations:
\begin{itemize}[leftmargin=*]
\item LLM-based methods generally outperform ID-based methods. On most datasets (Instruments, Movie, and Industry), the Top-N recommendation metrics of LLM-based methods consistently surpass those of ID-based methods. Similarly, on the Toys dataset, the majority of LLM-based methods lead their ID-based counterparts. This demonstrates that the superior understanding capabilities of LLMs enable the capture of richer semantic information, thereby enhancing Top-N recommendation performance.
\item Incorporating latent reasoning improves recommendation performance for both ID-based and LLM-based approaches. On the Instruments dataset, ReaRec, which introduces autoregressive prediction within the ID latent space, achieves improvements of 17.71\% and 15.08\% in H@5 and N@5, respectively, compared to SASRec. Similarly, $LaRec$ achieves improvements of 10.34\% and 9.16\% over TallRec. This indicates that performing latent reasoning effectively captures user preferences within the latent space, establishing more accurate connections between user history and target items.
\item The proposed $LaRec$ significantly outperforms all baselines across all evaluation metrics on four datasets. Furthermore, our approach achieves varying degrees of improvement compared to other reasoning-enhanced methods. This consistent performance superiority fully attests to its advanced capabilities in sequential generative recommendation tasks. 
% Specifically, compared to LatentR3, which directly introduces latent reasoning, $LaRec$ first enables the LLM to learn how to "think implicitly" within the latent space through Latent Pre-training, and subsequently guides the reasoning process to fully unleash its potential via Personalized Reinforcement Alignment.
\end{itemize}

\subsection{Ablation Study}

\begin{table}[h]
\centering
\caption{Ablation study of $LaRec$ across four datasets (H@10 and N@10).}
\label{ablation}
% \vspace{-3mm}
\resizebox{\columnwidth}{!}{
    \renewcommand{\arraystretch}{1.5}
    \setlength{\tabcolsep}{5pt} % 由于列数减少，可以适当增加列间距提高可读性
    
    \begin{tabular}{l|l|cc|cc|cc|cc}
    \toprule
    
    % === 表头 ===
    \multicolumn{2}{c|}{\multirow{2}{*}{\textbf{Variants}}} & \multicolumn{2}{c|}{\textbf{Toy}} & \multicolumn{2}{c|}{\textbf{Instruments}} & \multicolumn{2}{c|}{\textbf{Movies}} & \multicolumn{2}{c}{\textbf{Industry}} \\ 
    \cline{3-10} 
    
    \multicolumn{2}{c|}{} & H@10 & N@10 & H@10 & N@10 & H@10 & N@10 & H@10 & N@10 \\ 
    \noalign{\hrule height 0.8pt}
    
    % === Ours ===
    \rowcolor{green!10}
    \multicolumn{2}{c|}{\textbf{$LaRec$}} & \textbf{0.1073} & \textbf{0.0695} & \textbf{0.1238} & \textbf{0.0964} & \textbf{0.2256} & \textbf{0.1370} & \textbf{0.4090} & \textbf{0.2509} \\ 
    \hline
    
    % === Module 1 ===
    \multirow{2}{*}{w/o PRL} & w/o PGD & 0.1053 & 0.0664 & 0.1233 & 0.0950 & 0.2210 & 0.1340 & 0.4061 & 0.2474 \\
                             & w/o RL  & 0.1050 & 0.0662 & 0.1229 & 0.0949 & 0.2206 & 0.1329 & 0.4058 & 0.2470 \\ 
    \hline
    
    % === Module 2 ===
    \multirow{2}{*}{w/o LPT} & w/o PA  & 0.0995 & 0.0640 & 0.1190 & 0.0923 & 0.2182 & 0.1302 & 0.4025 & 0.2441 \\
                             & w/o SA  & 0.0948 & 0.0596 & 0.1142 & 0.0901 & 0.2161 & 0.1285 & 0.3948 & 0.2407 \\ 
    \hline
    
    % === Module 3 ===
    \multicolumn{2}{c|}{w/o LR}       & 0.0900 & 0.0566 & 0.1106 & 0.0893 & 0.2083 & 0.1270 & 0.3654 & 0.2174 \\
    
    \bottomrule
    \end{tabular}
}
\end{table}

We conduct ablation studies to validate the contributions of the core components within $LaRec$, specifically Latent Pre-training (LPT) (where w/o PA and w/o SA denote the removal of Process direction Alignment and Step-level Alignment, respectively), Personalized RL-tuning (PRL) (where w/o PGD and w/o RL denote the removal of the Personalized Guidance Distribution and Reinforcement Alignment, respectively), and Latent Reasoning (LR). As shown in Table \ref{ablation}, we progressively removed these components and derived the following experimental conclusions:
\begin{itemize}[leftmargin=*]
\item  Removing any single component (either Latent Pre-training or Personalized Reinforcement Alignment) leads to a noticeable decline in model performance across all datasets. The model achieves its worst performance when all components are removed. This indicates that both alignment objectives in the Latent Pre-training phase are essential core modules, while personalized RL-tuning provides additional performance gains. 
\item  The model variant integrating both Step-level Alignment and Process direction Alignment outperforms the version using Step-level Alignment alone. This corroborates that the two alignment strategies offer complementary advantages: Step-level Alignment transfers knowledge from explicit CoT to enhance reasoning coherence at the step level, while Process direction Alignment improves the correctness of the reasoning direction.
\item In the Personalized RL-tuning (PRL) phase, replacing the "Personalized Guide Distribution" with standard Gaussian noise (denoted as w/o PGD) results in recommendation performance significantly inferior to $LaRec$. This directly validates our "Anchored Exploration" hypothesis: in a sparse recommendation space, blind random perturbations struggle to effectively explore correct paths, whereas personalized exploration based on the neighborhood of user historical interests can efficiently discover paths leading to the correct answer.
\end{itemize}

\subsection{Further Analysis}
% \begin{figure*}[h]
% \includegraphics[width=0.99 \textwidth]{images/further_analysis.pdf}
% \vspace{-3mm}
% \caption{Further Analysis of $LaRec$.}
% \label{fig:effect}
% \end{figure*}

\subsubsection{Analysis of Reasoning Process Effectiveness}
\begin{figure}[h]
\includegraphics[width=0.4 \textwidth]{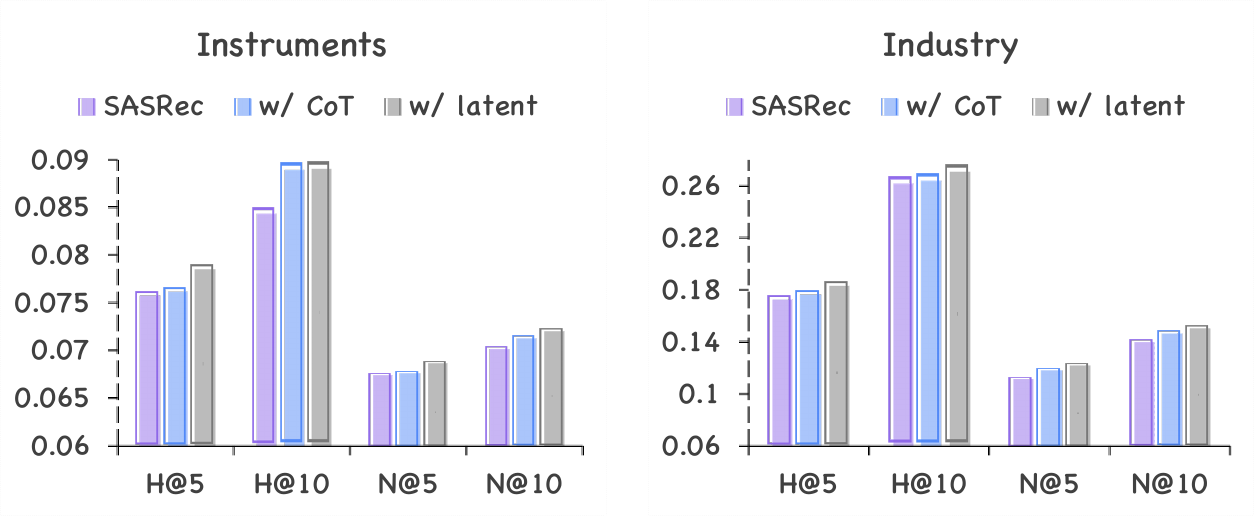}
% \vspace{-3mm}
\caption{Analysis of Reasoning Process Effectiveness.}
\label{fig:effect}
\end{figure}

To validate the effectiveness of latent reasoning in $LaRec$, we select the classic SASRec model as our evaluation backbone. We inject both the hidden states from Explicit CoT and the latent states from implicit reasoning into SASRec as user embeddings to observe their respective gains in recommendation performance. Specifically, we utilize the hidden state corresponding to the final token of the explicit CoT as the standard representation of explicit user preference. Conversely, we employ the average pooling of the $K$-step latent reasoning states as the standard representation of implicit user preference. As shown in Figure \ref{fig:effect}, we observe that injecting reasoning-based user representations into SASRec consistently enhances recommendation performance. Notably, the implicit user representations achieve results that are comparable to, or even surpass, those of the explicit user representations. This demonstrates that reasoning representations derived from Latent Pre-training can more precisely capture complex user preferences, thereby validating the effectiveness of the dual alignment strategy within Latent Pre-training.

\subsubsection{Impact of Reasoning Length.}
\begin{figure}[h]
% \vspace{-3mm}
\includegraphics[width=0.4\textwidth]{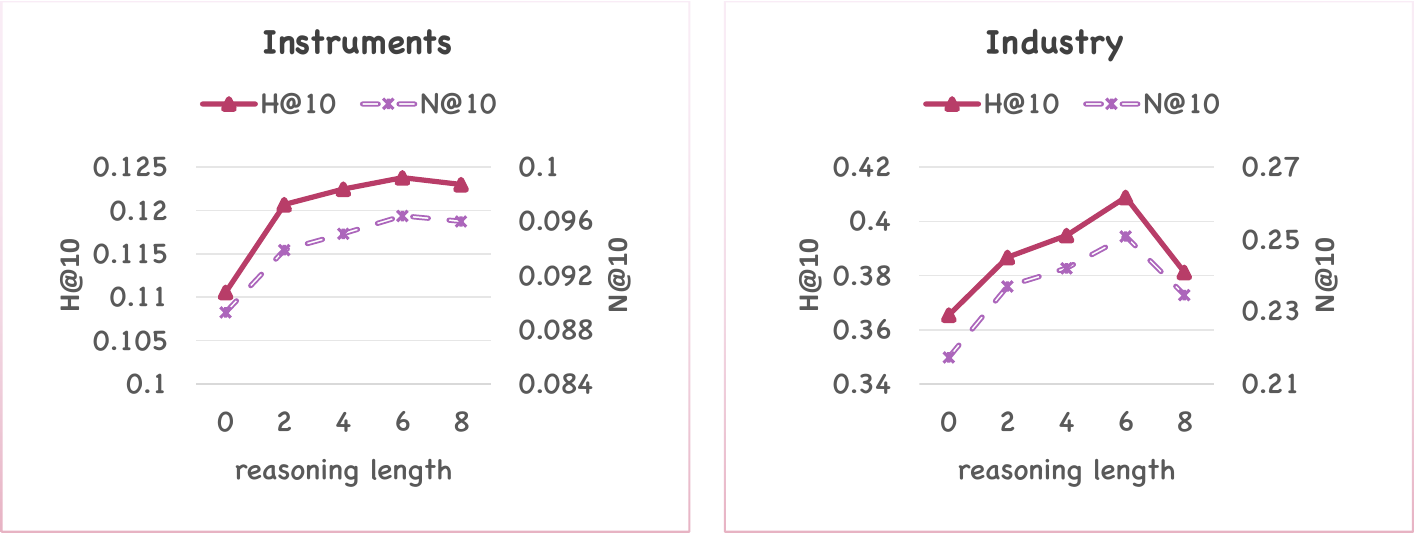}
% \vspace{-3mm}
\caption{Impact of Reasoning Length.}
\label{fig:length}
% \vspace{-5mm}
\end{figure}

\begin{figure}[h]
% \vspace{-3mm}
\includegraphics[width=0.4\textwidth]{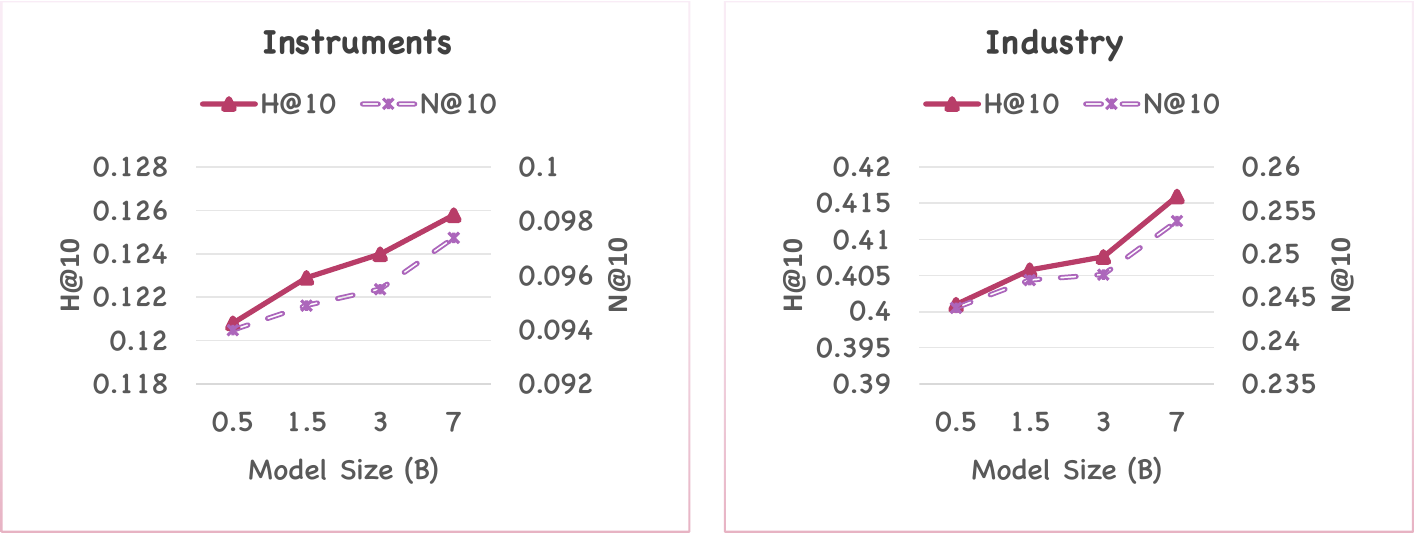}
% \vspace{-3mm}
\caption{Impact of LLM Scaling Laws.}
\label{fig:scale}
% \vspace{-3mm}
\end{figure}

\begin{figure*}[t]
\includegraphics[width=0.95\textwidth]{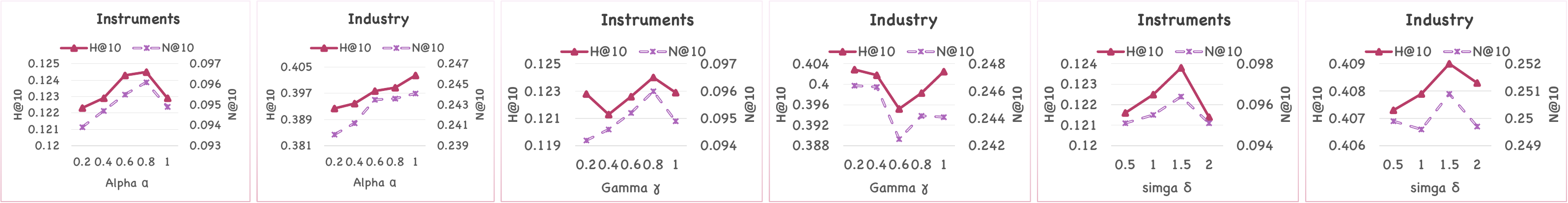}
% \vspace{-3mm}
\caption{Impact of different alignment coefficients $\alpha$ , $\gamma$ and $\sigma$.}
\label{fig:hyper}
% \vspace{-3mm}
\end{figure*}
To analyze the impact of the number of latent reasoning steps, we vary the reasoning steps $K$ within the set $\{0, 2, 4, 6, 8\}$ (where $0$ indicates no reasoning). The experimental results in Figure \ref{fig:length} exhibit a consistent trend: as the number of reasoning steps increases, recommendation performance initially improves, reaches an optimal peak, and subsequently begins to decline. We observe that the metrics for reasoning instances ($K \in \{2, 4, 6, 8\}$) consistently outperform those without reasoning ($K=0$), thereby demonstrating the effectiveness of the implicit reasoning mechanism. This observation aligns with findings from previous studies such as ReaRec\cite{rearec} and LARES \cite{lares}. Furthermore, we find that both datasets achieve optimal performance at $K=6$, which corresponds to the number of reasoning steps present in the majority of samples within our constructed data for Step-level Alignment. We attribute this to the fact that during Latent Pre-training, too few steps may fail to fully transfer the semantic logic of the Explicit CoT, leading to insufficient user reasoning; conversely, excessive steps may result in "over-thinking". Therefore, to maximize the efficacy of implicit reasoning, it is crucial to select an appropriate number of steps during the training phase.

\subsubsection{Impact of LLM Scaling Laws.}
To verify whether latent reasoning exhibits scaling laws analogous to explicit reasoning, we implemented $LaRec$ using the Qwen2.5 series LLMs across varying parameter scales on the Instruments and Industry datasets. As illustrated in Figure \ref{fig:scale}, the recommendation performance of $LaRec$ consistently improves as the model size increases. This demonstrates that implicit reasoning also adheres to the scaling laws characteristic of LLMs.

\subsubsection{Hyperparameter Sensitivity}
% \begin{figure}[h]
% \includegraphics[width=0.4\textwidth]{images/hyper.pdf}
% \vspace{-3mm}
% \caption{Impact of different alignment coefficients $\alpha$ and $\gamma$.}
% \label{fig:hyper}
% \end{figure}
We investigate the impact of the step-level alignment coefficient $\alpha$ and process direction alignment coefficient $\gamma$. 
% We evaluate model performance on the Instruments and Industry datasets across different combinations of the hyperparameter $\alpha$ (with values in $\{0.2, 0.4, 0.6, 0.8, 1.0\}$) and $\gamma$ (with values in $\{0.2, 0.4, 0.6, 0.8, 1.0\}$). 
As illustrated in Figure \ref{fig:hyper}, the performance of $LaRec$ on both datasets exhibits fluctuations as $\alpha$ increases, achieving optimal results at $\alpha=0.8$ and $\alpha=1.0$, respectively. 
% This suggests that step-level alignment exerts a positive influence on model learning, indicating that a larger coefficient is preferable during training. 
This suggests that a larger step-level alignment weight generally facilitates model learning.
% Regarding the hyperparameter $\gamma$, performance on the Instruments dataset initially improves as the value increases but declines within the higher value range. Conversely, on the Industry dataset, performance generally decreases at first before exhibiting a steady upward trend. These findings underscore the critical role of these hyperparameters. 
Regarding $\gamma$, the performance follows an initial increase followed by a decline on Instruments, whereas on the Industry dataset, it exhibits a steady upward trend after an initial dip. These variations reflect the differing sensitivities of various data distributions to process direction alignment.
Furthermore, we conduct an analysis of the local exploration radius $\sigma$ in personalized RL-tuning. The performance of $LaRec$ exhibits an initial increase followed by a decline as the value of $\sigma$ increases, reaching its peak effectiveness at $\sigma = 1.5$. This indicates that an appropriate exploration space facilitates reinforced alignment for the LLM, thereby enhancing recommendation performance.

\subsubsection{Analysis of Reasoning Efficiency.}
% \begin{figure}[h]
% \vspace{-5mm}
% \includegraphics[width=0.3\textwidth]{images/efficiency_v2.pdf}
% \vspace{-3mm}
% \caption{Analysis of Reasoning Efficiency.}
% \label{fig:efficiency}
% \vspace{-3mm}
% \end{figure}

\begin{table}[htbp]
% \vspace{-3mm}
\centering
\caption{Inference Latency Comparison across Different Datasets (seconds per sample).}
\label{tab:latency_results}
% \vspace{-3mm}
\begin{tabular}{lcccc}
\toprule
\textbf{Method} & \textbf{Toys} & \textbf{Instruments} & \textbf{MovieLens} & \textbf{Industry} \\ \midrule
TALLRec         & 0.20s         & 0.23s                & 0.16s              & 0.56s             \\
Ours            & 0.24s         & 0.34s                & 0.19s              & 0.67s             \\
CoT             & 2.49s         & 2.88s                & 2.85s              & 3.92s             \\ \bottomrule
\end{tabular}
% \vspace{-3mm}
\end{table}
To systematically evaluate the inference efficiency of $LaRec$, we conduct efficiency tests across all four experimental datasets. We compare the per-sample inference latency of the method under three conditions: without a reasoning mechanism (w/o Reason), with latent reasoning ($K=6$), and with Explicit CoT reasoning. All tests were executed on a single H20 GPU. As illustrated in Table \ref{tab:latency_results}, the latency of our implicit reasoning approach is nearly on par with that of the non-reasoning method and is significantly lower than the time required for explicit reasoning mechanisms. These results demonstrate that $LaRec$ effectively achieves an optimal balance between effectiveness and efficiency, maintaining low latency while ensuring high recommendation performance.

\subsubsection{Analysis of User Preference Strategy.}
% We further analyze our preference reasoning strategies. By comparing Non-Target CoT Reasoning with Target-Oriented CoT (as shown in Table \ref{tab:prompt_analysis}), we observe that the superior performance of Target-Oriented CoT stems from the fact that a clear reasoning objective generates higher-quality trajectories, rather than from information leakage.
We further analyze the user preference reasoning strategy. By comparing Non-Target CoT with Target-Oriented CoT (as shown in Table \ref{tab:prompt_analysis}), we observed that Target-Oriented CoT achieves better performance, indicating that target-oriented CoT reasoning can generate higher-quality user preference trajectories without information leakage.
\begin{table}[h]
\centering
\caption{Analysis of reasoning strategies.}
\label{tab:prompt_analysis}
\vspace{-2mm}
\resizebox{0.95\columnwidth}{!}{
\begin{tabular}{l|cc|cc}
\toprule
\multirow{2}{*}{\textbf{Reasoning Strategy}} & \multicolumn{2}{c|}{\textbf{Instruments}} & \multicolumn{2}{c}{\textbf{Industry}} \\
\cline{2-5}
 & H@10 & N@10 & H@10 & N@10 \\
\midrule
Non-Target CoT & 0.1225 & 0.0950 & 0.4071 & 0.2485 \\
\rowcolor{green!10} 
\textbf{Target-Oriented CoT} & \textbf{0.1238} & \textbf{0.0963} & \textbf{0.4090} & \textbf{0.2509} \\
\bottomrule
\end{tabular}
}
\end{table}

\begin{table}[h]
% \vspace{-3mm}
\centering
\caption{Online A/B Test Results.}
\vspace{-3mm}
\begin{tabular}{lccc}
\toprule
&\textbf{Exposure} & \textbf{Costs} & \textbf{Conversion}\\ \hline
$LaRec$        & +0.46\%  &  +1.39\%  & +2.93\% \\ \bottomrule
\end{tabular}
\label{online}
% \vspace{-6mm}
\end{table}

\subsection{Online Performance}
We deployed the proposed $LaRec$ on a real-world advertising platform with hundreds of millions of active users as an additional LLM-based Recall channel. We use the text description of the items that the user clicked on in two months to construct the prompts and export the output embeddings as key vector to retrieve similar advertisements by Approximate Nearest Neighbor~(ANN). The item embeddings are generated by LLM with the content and keywords of advertisements. All embeddings are updated daily by $LaRec$. The comparison baseline is the online system that contains multiple ID-based and text-based Recall channels. We select the number of \textbf{Exposure}, the number of \textbf{Conversion}, and \textbf{Costs} as the primary metrics. And the A/B test is conducted for a weak to get credible performance.
The results, shown in Table~\ref{online}, reveal that $LaRec$ achieved significant improvements in all metrics, highlighting its strong potential for real-world deployment in the industrial recommender system.

\section{Conclusions}
% We propose $LaRec$, an efficient generative recommendation framework. $LaRec$ performs recommendations via latent reasoning, simultaneously enhancing recommendation performance and improving inference efficiency. To fully unleash latent reasoning capabilities of LLMs, we design a two-stage optimization strategy. First, $LaRec$ introduces Latent Pre-training mechanism. By constructing step-level alignment and process direction alignment, Latent Pre-training provides dense supervisory signals for the evolution of each latent state, effectively mitigating the issue of process supervision sparsity. Second, to overcome the constraints of deterministic reasoning and avoid blind exploration, we propose a Personalized RL-tuning. This approach further unlocks the model's reasoning potential by incentivizing it to efficiently explore diverse reasoning paths. Extensive experiments demonstrate that $LaRec$ achieves superior recommendation performance while maintaining exceptional inference efficiency.
We propose $LaRec$, an efficient generative recommendation framework that enhances performance through latent reasoning while maintaining low inference latency. To fully unleash the reasoning potential of LLMs, we introduce a two-stage optimization strategy: (1) Latent Pre-training, which provides dense supervisory signals via step-level and process-direction alignment to mitigate process sparsity; and (2) Personalized RL-tuning, which encourages the exploration of diverse, optimal reasoning paths while avoiding deterministic constraints. Extensive experiments demonstrate that $LaRec$ achieves superior recommendation performance while maintaining exceptional inference efficiency.

\bibliographystyle{ACM-Reference-Format}
\balance
\bibliography{main}

\end{document}